\documentclass[aps,pre,preprint,superscriptaddress]{revtex4}
\usepackage[dvipdfmx]{graphicx}
\usepackage{ulem}
\usepackage{color}
\def\vector#1{\mbox{\boldmath $#1$}}

\begin{document}

\title{Temperature-field phase diagram of the two-dimensional dipolar Ising ferromagnet}

\author{Hisato \surname{Komatsu}}
\email[Email address: ]{komatsu.hisato@nims.go.jp}

\affiliation{Research Center for Advanced Measurement and Characterization, National Institute for Materials Science, Tsukuba, Ibaraki 305-0047, Japan}
\affiliation{International Center for Materials Nanoarchitectonics,  National Institute for Materials Science,  Tsukuba, Ibaraki 305-0044, Japan}

\author{Yoshihiko Nonomura}
\email[Email address: ]{nonomura.yoshihiko@nims.go.jp}

\affiliation{International Center for Materials Nanoarchitectonics,  National Institute for Materials Science,  Tsukuba, Ibaraki 305-0044, Japan}

\author{Masamichi Nishino}
\email[Email address: ]{nishino.masamichi@nims.go.jp} 

\affiliation{Research Center for Advanced Measurement and Characterization, National Institute for Materials Science, Tsukuba, Ibaraki 305-0047, Japan}
\affiliation{International Center for Materials Nanoarchitectonics,  National Institute for Materials Science,  Tsukuba, Ibaraki 305-0044, Japan}

\begin{abstract}

We study field-induced phase transitions in the two-dimensional dipolar Ising ferromagnet with a specific ratio between the exchange and dipolar constants, $\delta=1$, which exhibits 
a stripe-ordered phase with the width of one lattice unit at low temperatures without magnetic field. By using a mean-field (MF) approximation and a Monte Caro (MC) method with the stochastic-cutoff algorithm, which is an $O(N)$ simulation method, we show the temperature-field phase diagrams. 
In the MC study the orientational order and the structure factor are evaluated. Second-order transition points are determined by a finite-size-scaling analysis  
and first-order transition points are identified by the analysis of the energy histogram. Although both the MF and MC phase diagrams consist of wide regions of several stripe-ordered phases and narrow regions between them characterized by complicated stripe patterns, they show qualitative and quantitative differences in possible phases and phase boundaries. In the MF phase diagram, three main stripe-ordered phases exhibit a nesting structure, 
while in the MC phase diagram, 
two main stripe-ordered phases 
are located separately, which causes a characteristic field-induced reentrant transition of the orientational order. 
\end{abstract}

\maketitle

\section{Introduction \label{Introduction} }

Thin magnetic films have attracted much attention 
in their possible applications such as magnetic recording~\cite{Bader}.   
They show a variety of ordering processes with stripe states, reorientation transitions, etc~\cite{ASB90,Kashuba,DeBell,Portmann}. 
The origin of the complexity of the phenomena is the competition between short-range exchange and long-range dipolar interactions. 

The two{\color{red}-}dimensional (2D) dipolar Ising ferromagnet has been intensively studied to catch the essence of the phenomena in the strong anisotropy limit. 
The zero-field phase diagram as a function of temperature and the ratio $\delta$ between the exchange and dipolar constants (see eq.~(\ref{Hamiltonian})) has been focused on~\cite{Booth,MacIsaac,Toloza,Rastelli06,Cannas,Pighin,Rastelli,Rizzi,Vindigni,Fonseca,Ruger,Horowitz,Bab}{\color{red}.} 
The dipolar interaction leads to the antiferromagnetic (AF) ground state~\cite{MacIsaac}, while the Ising ferromagnetic exchange interaction causes the ferromagnetic ground state. 
Hence frustration affects the ordering process when the two interactions coexist. 

MacIsaac et al. reported~\cite{MacIsaac} that the ground state changes at $\delta=0.425$ from the AF state to a stripe-ordered state, 
in which neighboring stripes have opposite magnetizations along the $z$ axis. 
On the other hand, Rastelli et al. pointed out~\cite{Rastelli} that the model exhibits irregular checkerboard configurations for $0.4152<\delta < 0.4403$ between the AF and stripe-ordered ones. 
They also presented the region of $\delta$ for the stability of stripe-ordered phases, e.g., the stripe-ordered phase with the stripe width of one lattice unit, $h=1$, is stable for $0.4403<\delta<1.2585$. 
The width of the stripe becomes larger with increasing $\delta$~\cite{MacIsaac,Rastelli}. 

At finite temperatures phase transitions between the stripe phases and the disordered (D) phase (or tetragonal liquid (TL) phase) occur, in which  $\pi/2$-rotational symmetry is broken. 
It has been clarified that the phase transitions between the stripe-ordered phase with the width $h\ge2$ and the D (TL) phase are mainly of first order~\cite{Rastelli06,Pighin,Rastelli}, and the nematic phase~\cite{Abanov} partially exists between the two phases. The phase transition between the stripe-ordered phase with $h=1$ and the D (TL) phase is still open to debate, namely the phase boundary is a second-order line or the second-order and first-order lines merge at some $\delta$ on the phase boundary~\cite{Pighin,Rastelli,Cannas,Fonseca,Horowitz,Bab}. 

The 2D dipolar Ising ferromagnet also shows field-induced phase transitions~\cite{Garel,Arlett}. 
D\'{i}az-Mendez and Mulet investigated the field-temperature phase diagram for 
$\delta=2$~\cite{MM10}, which gives the alternating stripes of the width of two lattice units at low temperatures without magnetic field. 
They presented perfect stripes, anharmonic stripes, bubbles, and ferromagnetic phases with increasing the field at low temperatures. 
They also pointed out that the phase transition between the bubble and ferromagnetic phases looks like the Berezinskii-Kosterliz-Thouless (BKT) one~\cite{Berezinskii,KT}.

In the present study we focus on the model (\ref{Hamiltonian}) with $\delta=1$, in which the ground state configuration without magnetic field is the perfect alternating stripes with the width of $h=1$. 
We investigate field-induced phase transitions but do not concentrate on the identification of the bubble phase at present because the correlation length shows an exponential divergence and furthermore logarithmic corrections are accompanied in the BKT phase transition, and precise detection of the BKT transition generally requires very large-scale simulations. 

Naive MC algorithms require $O(N^2)$ computational time for simulation of long-range interaction systems such as dipolar systems, since the number of the interactions is $O(N^2)$. This causes difficulty in the simulation of large system sizes. To overcome this difficulty, several $O(N)$ MC algorithms have been proposed~\cite{SM08,FT09,H18}. 
In the present study we adopt the stochastic cutoff (SCO) method proposed by Sasaki and Matsubara~\cite{SM08}. 
This $O(N)$ algorithm was first introduced to a Heisenberg dipolar system, in which the SCO procedure was applied to all dipolar interactions.   
To realize an efficient MC sampling in the dipolar Ising system with the SCO algorithm, we tune the range (number) of the interactions to which this algorithm is applied.

The rest of the paper is organized as follows. 
The model is presented in Sec.~\ref{model}. 
A mean-field (MF) approximation is performed and the MF phase diagram is given 
in Sec.~\ref{MF}. 
In Sec.~\ref{simulation} the phase diagram is studied by the $O(N)$ MC simulation. 
Sec.~\ref{summary} is devoted to the summary of the paper including 
the comparison between the MF and MC phase diagrams. 
In Appendix A, a tricritical point in the MF phase diagram is derived. 
In Appendix B, the SCO method and the realization of 
efficient sampling for the dipolar Ising system with the use of the SCO method are briefly explained.

\section{Model}
\label{model}

The Hamiltonian of the 2D dipolar Ising ferromagnet is 
\begin{equation}
{\cal H} = - \delta \sum _{\left< i,j \right> } \sigma _i \sigma _j + \sum _{i < j  } \frac{\sigma _i \sigma _j }{r_{ij} ^3} - H \sum _i \sigma _i. 
\label{Hamiltonian}
\end{equation}
The spin variable $\sigma _i$ takes $\sigma _i=1$ (up) or $\sigma _i=-1$ (down), perpendicular to the 2D plane.  
Here $\delta (>0)$ is the ratio between the exchange and dipolar constants, and $H$ is the magnetic field. 
The distance between sites $i$ and $j$, $r_{ij}$, is measured in units of the lattice constant. 
The first sum $\left< i,j \right>$ runs over all pairs of nearest-neighbor spins and the second one over all pairs of spins. 
We focus on square-lattice systems with $N=L \times L$ sites.

\section{Mean-field analysis \label{MF} }

First we investigate the phase diagram of the system (\ref{Hamiltonian}) using a MF approximation. 
Here we assume that the magnetization is uniform in each column in the system of $L$ columns and stripes have a period of $n$ columns~\cite{Vindigni,Pighin}. The self-consistent equations for the system (\ref{Hamiltonian}) are given by  
\begin{equation}
m_k = \tanh \left\{ \beta \left( \sum _{l =1} ^{n} \tilde{U} _{k,l} m_l + H \right) \right\},
\label{SCeq}
\end{equation}
where $\beta$ is the inverse temperature, i.e., $\beta=1/(k_{\rm B}T)$.  
We set $k_{\rm B}=1$ from now on. 
Here $m_k(=m_{k+n})$ is the average per-site magnetization at the $k$-th row ($k=1,2,\cdots n$) and \begin{equation}
 \ \ \tilde{U} _{k,l} =\sum _{\alpha=-\infty} ^{\infty} U_{k, l+ \alpha n }, 
\end{equation}
where $U_{k,l}$ is defined as 
 \begin{equation}
U_{k,l}  =  \delta \left(  2 \delta _{k,l} + \delta _{k,l+1} + \delta _{k,l-1} \right)-\sum _{\nu =-\infty} ^{\infty} \frac{1}{ \left| (l-k) ^2 + \nu ^2 \right| ^{3/2}   }, 
\end{equation}
with integers $\alpha$ and $\nu$. 
%
\begin{figure}[b]
\includegraphics[width = 4cm]{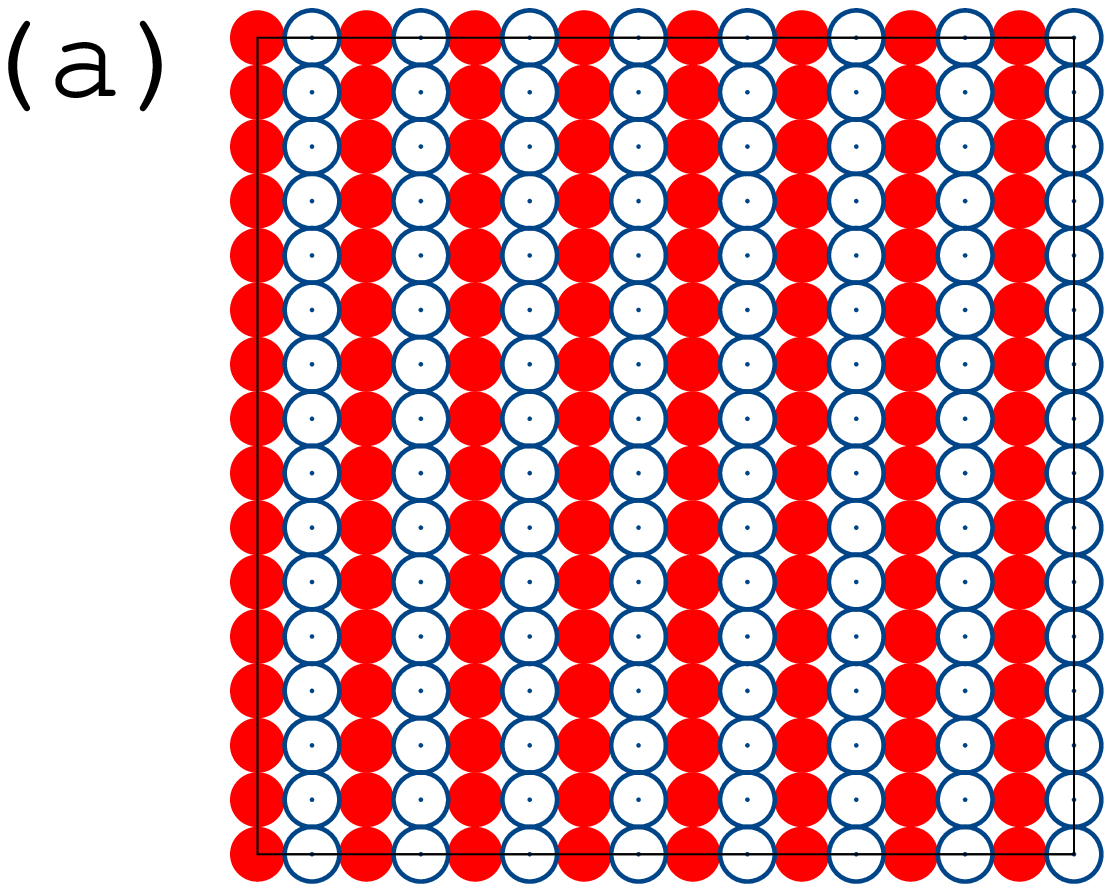}
\includegraphics[width = 4cm]{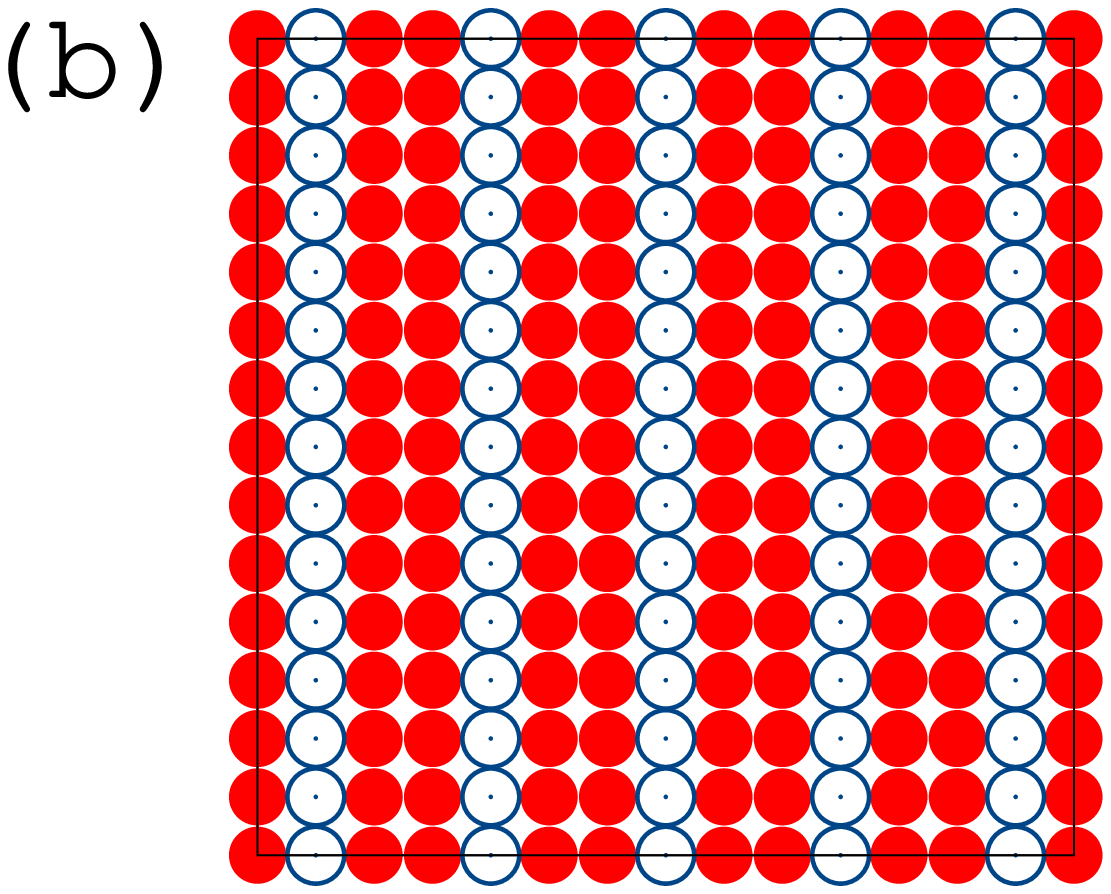}
\includegraphics[width = 4cm]{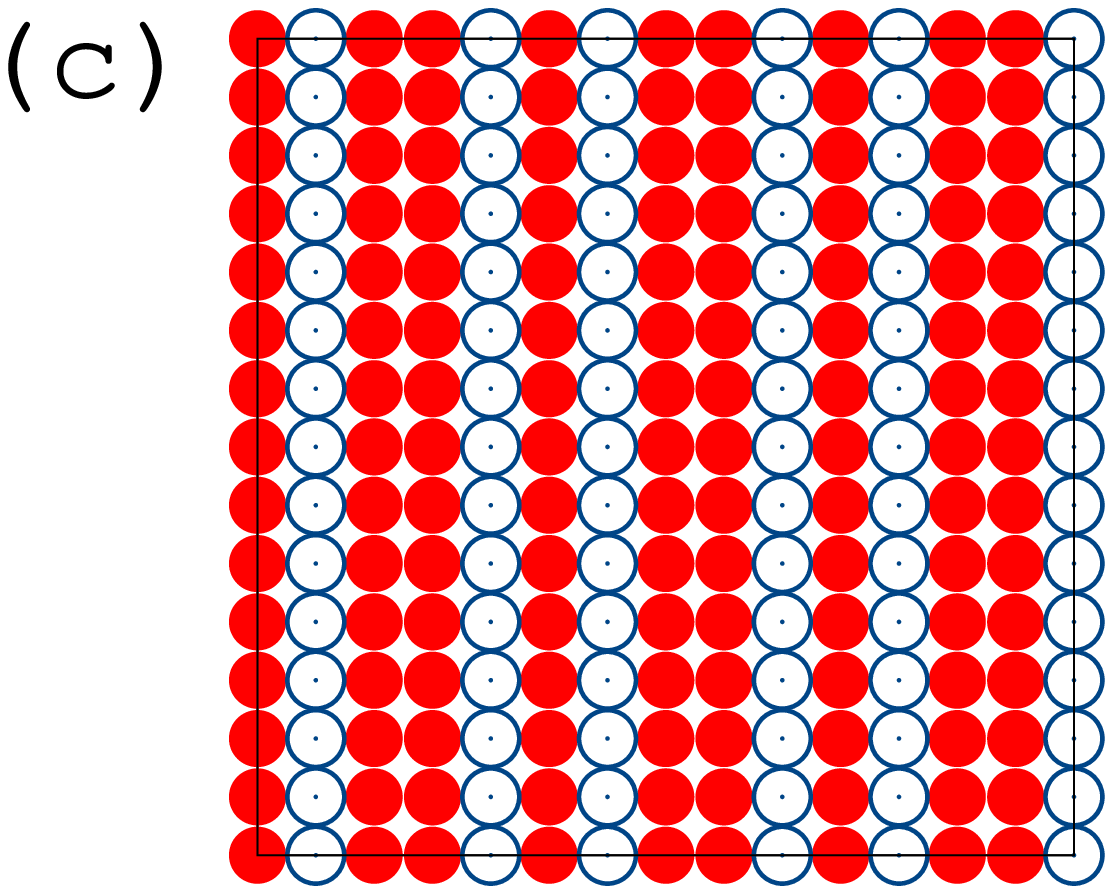}
\caption{Examples of stripe configurations in the ground states in the 
(a) $\langle 1 \rangle$ phase, (b) $\langle 21 \rangle$ phase, and (c) $\langle 21^3 \rangle$ phase. The red filled and blue open circles denote up and down spins, respectively.} 
\label{Config}
\end{figure}

We classify stripe phases by using the notation introduced in Ref.~\cite{Pighin}: $\langle h_1^{n_1} h_2^{n_2} \cdots h_m^{n_m}  \rangle$, in which $h_i$ is the width of a stripe and $n_i$ is the number of consecutive stripes with the same width $h_i$. We consider a stripe as a region which has a uniform magnetization. Examples of the ground-state configurations of stripe phases are given in Fig.~\ref{Config}. It should be noted that a phase with 1-row stripe is written as $\langle 1 \rangle$ (Fig.~\ref{Config} (a)). 
In the present study we investigate stripe phases by evaluating the free energies for the solutions of the self-consistent field (SCF) equations (\ref{SCeq}).

We solve the SCF equations (\ref{SCeq}) numerically. We calculate up to $n=8$ in most cases and to $n=11$ for a part of complicated phases (see below). In the summations about $\alpha$ and $\nu$ in eqs. (3) and (4), we calculate up to $|\alpha|=[10000/n]$ and $|\nu|=10000$. We determine stripe phases by comparing the stabilities of the free energies for the stable solutions:
\begin{eqnarray}
F & =  & \frac{L^2}{n}  \left[ \frac{1}{2} \sum _{i,j=1} ^n \tilde{U} _{i,j} m_i m_j  \right. \nonumber \\
& & \left. - \beta ^{-1} \sum _i \log \left[ 2 \cosh \left\{ \beta \left( \sum _{j=1} ^n \tilde{U} _{i,j} m_j + H \right) \right\} \right]   \right] .
\label{F_MF}
\end{eqnarray}

\begin{figure}[t!]
\begin{center}
\includegraphics[width = 7.0cm]{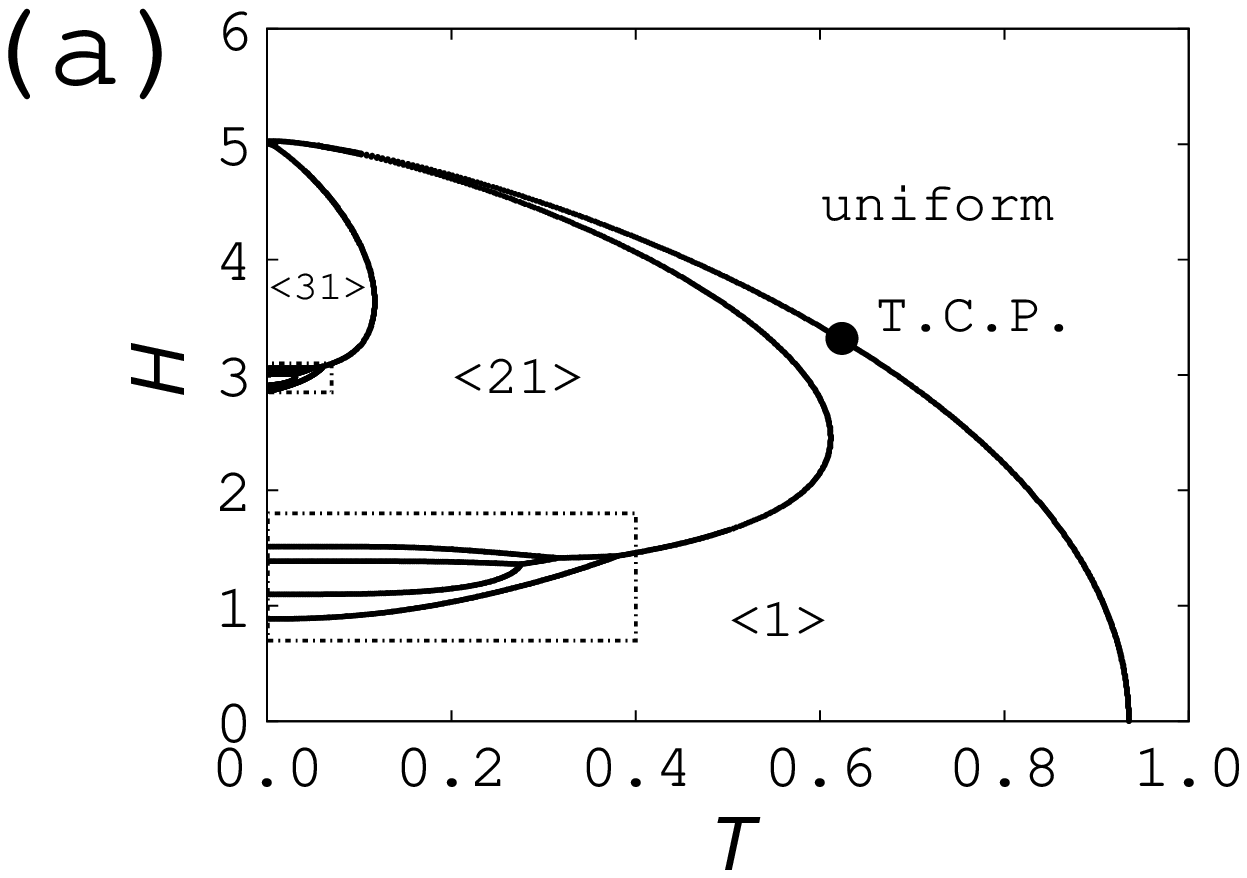}
\end{center}
\begin{center}
\includegraphics[width = 7.0cm]{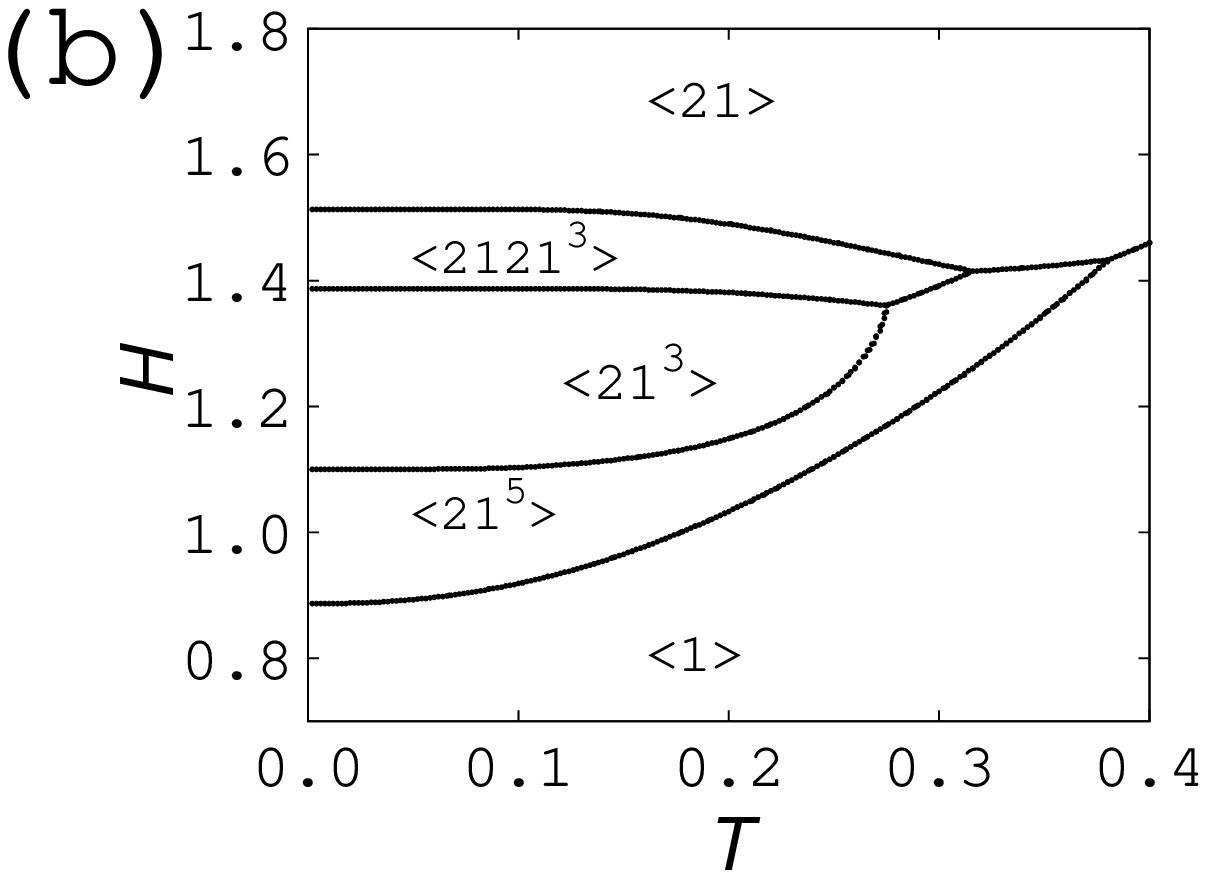}
\includegraphics[width = 7.0cm]{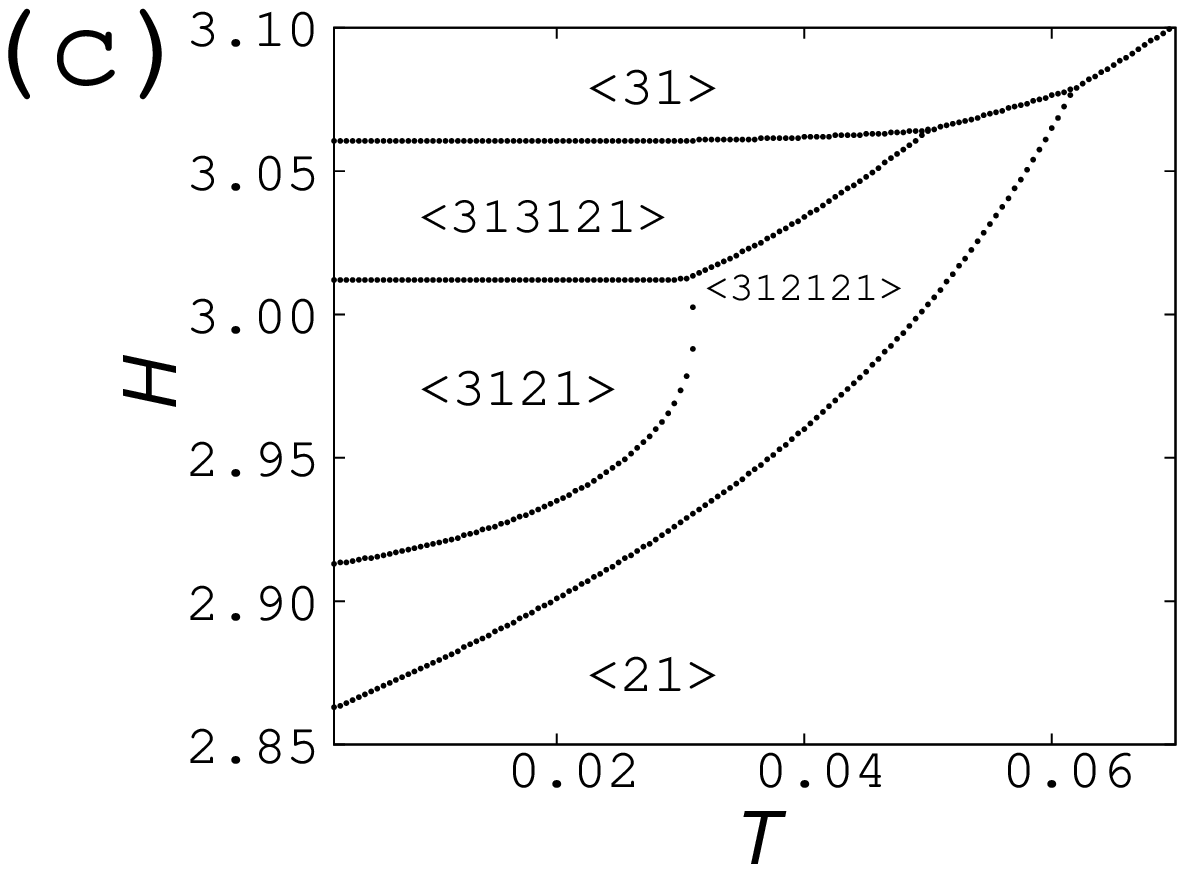} 
\end{center}
\caption{Mean-field phase diagram (a) in full scale with the tricritical point denoted by a circle, and its magnification of the (b) lower and (c) upper dotted frames.
}
\label{PD_MF1}
\end{figure}

Details of the MF phase diagram are given in Figs.~\ref{PD_MF1} (a)-(c). 
The system is ferromagnetic at high fields and low temperatures, 
while it is disordered at low fields and high temperatures. 
In the both states the orientational order is broken and they are not distinguishable. Here we call them ``uniform phase". 

In Fig.~\ref{PD_MF1} (a), $\langle1\rangle$, $\langle21\rangle$, and $\langle31\rangle$ phases are stable in wide regions.
We find that the transition between the $\langle1\rangle$ and uniform phases is of second order at low fields ($H < 3.316$), while it is of first order at high fields ($H > 3.316$), 
and there exists a tricritical point at $(T,H) \simeq (0.6236, 3.316)$ 
(see  Appendix~\ref{App1}). 
The other phase transitions are of first order. 

Complicated stripe phases, that is, the $\langle21^3\rangle$, $\langle2121^3\rangle$, and $\langle21^5\rangle$ phases between the $\langle1\rangle$ and $\langle21\rangle$  phases (Fig.~\ref{PD_MF1} (b)) are stable at low temperatures within the analysis up to $n=8$, and other complicated phases, i.e., the $\langle3121\rangle$, $\langle312121\rangle$, and $\langle313121\rangle$ phases between the $\langle21\rangle$ and $\langle31\rangle$ phases (Fig.~\ref{PD_MF1} (c)) are stable up to $n=11$.

\section{Monte Carlo study}
\label{simulation}

The MF approach neglects thermal fluctuation effects, 
and it generally leads to the overestimation of transition temperatures and fields and often fails to estimate the type of phase transitions, e.g., first order or second order. 
In this section we study the phase diagram of the system (\ref{Hamiltonian}) by $O(N)$ MC simulations based on the SCO algorithm~\cite{SM08}. For efficient MC sampling for the dipolar Ising system, we tune the range of the dipole interactions to which the SCO is applied (see Appendix~\ref{AppA}). 

\begin{figure}[!hbp]
\includegraphics[width = 11.5cm]{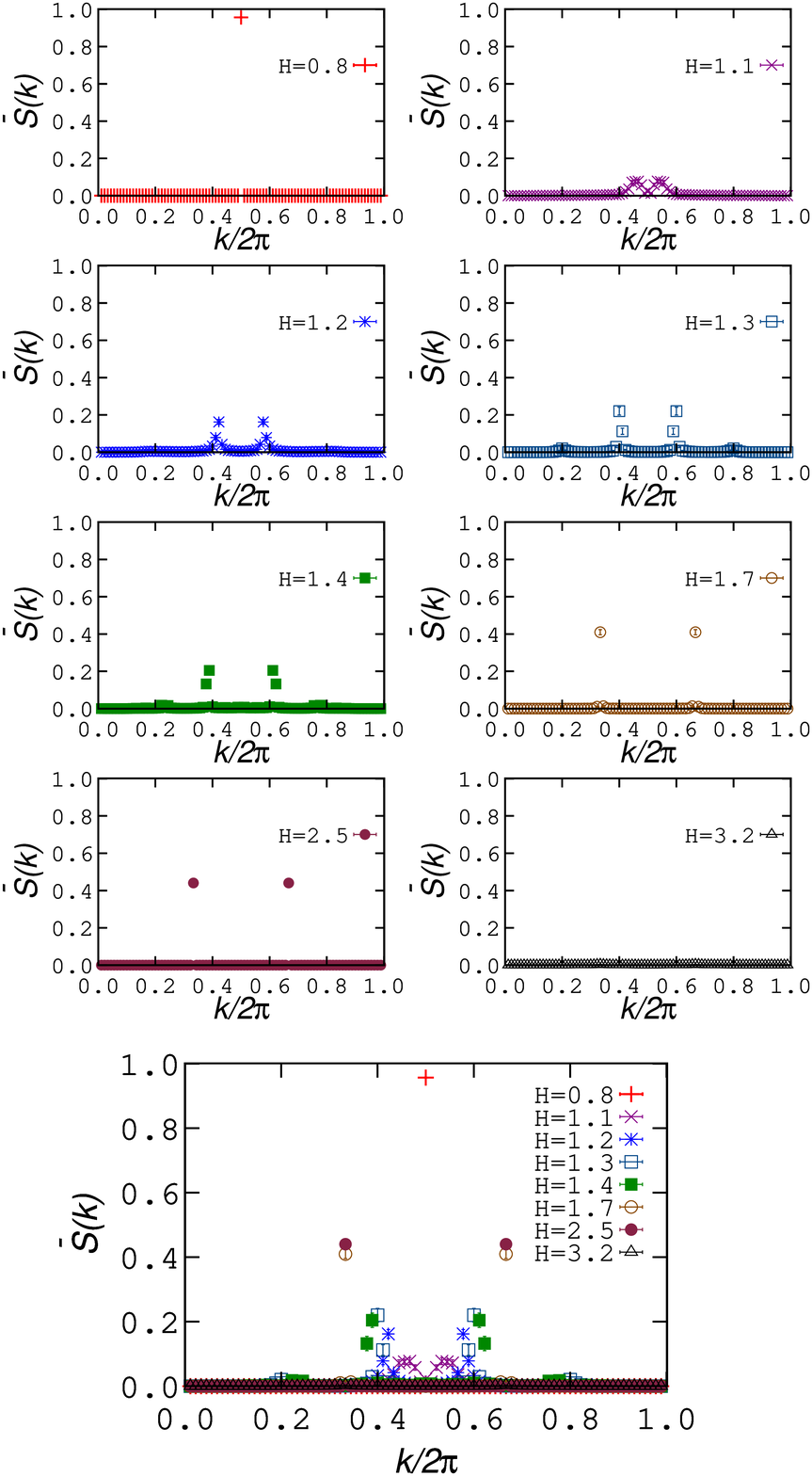}
\caption{Field dependence of $\bar{S} (k)$ at $T = 0.1$ for $L=90$ for 
H=0.8 (crosses), 1.1 (X-marks), 1.2 (stars), 1.3 (open squares), 1.4 (filled squares), 
1.7 (open circles), 2.5 (filled circles), and 3.2 (triangles). }
\label{Sk}
\end{figure}

\begin{figure}[!tbp]
\begin{center}
\includegraphics[width = 7.0cm]{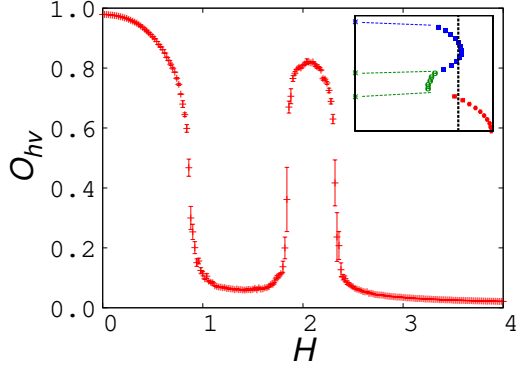}
\end{center}
\caption{Field dependence of $O_{hv}$ at $T=0.3$ for $L=48$. 
Location of this field sweep in the MC phase diagram is displayed in the inset with the dashed line.
}
\label{Field-dep}
\end{figure}

We take $2 \times 10^5$ Monte Caro steps (MCS) for equilibration and $3 \times 10^5$ MCS for measurement at each temperature (or magnetic field) in gradual change of temperature (or magnetic field) staring from random initial configurations. For Figs.\ \ref{Sk} and \ref{Field-dep}, we take $1.6 \times 10^{5}$ MCS for equilibration and $0.4 \times 10^{5}$ MCS for measurement. We obtain physical quantities by averaging over independent 48 measurements with different random number sequences. For Fig.~\ref{Field-dep} the average physical quantity is estimated by independent 4 measurements.

In order to exclude the effect of edges in the model with long-range interactions (\ref{Hamiltonian}),   we adopt a replica method tiling replicas of the original system~\cite{Ruger,Horowitz} with periodic boundary conditions. Here we tile 2001$\times$2001 replicas.

To detect stripe phases, we define the following order parameter~\cite{Booth,Cannas},
\begin{equation}
 O_{hv} \equiv \left| \frac{n_v - n_h}{n_v + n_h} \right| \label{Ohv}, 
\end{equation}
where $n_v$ and $n_h$ are the numbers of vertical and horizontal bonds between nearest neighbor anti-aligned spins, respectively. This order parameter detects $\pi/2$-rotational symmetry breaking, and if each row or column is fully ordered alternately, i.e., alternation of $\langle \sigma_i \rangle=1$ and $-1$, we have $\langle O_{hv} \rangle =1$. Here $\langle X  \rangle$ stands for the statistical average of the quantity $X$. For ferromagnetically-ordered or disordered states, we have $\langle O_{hv} \rangle = 0$.

\begin{figure}[tp]
\begin{center}
\includegraphics[width = 7cm]{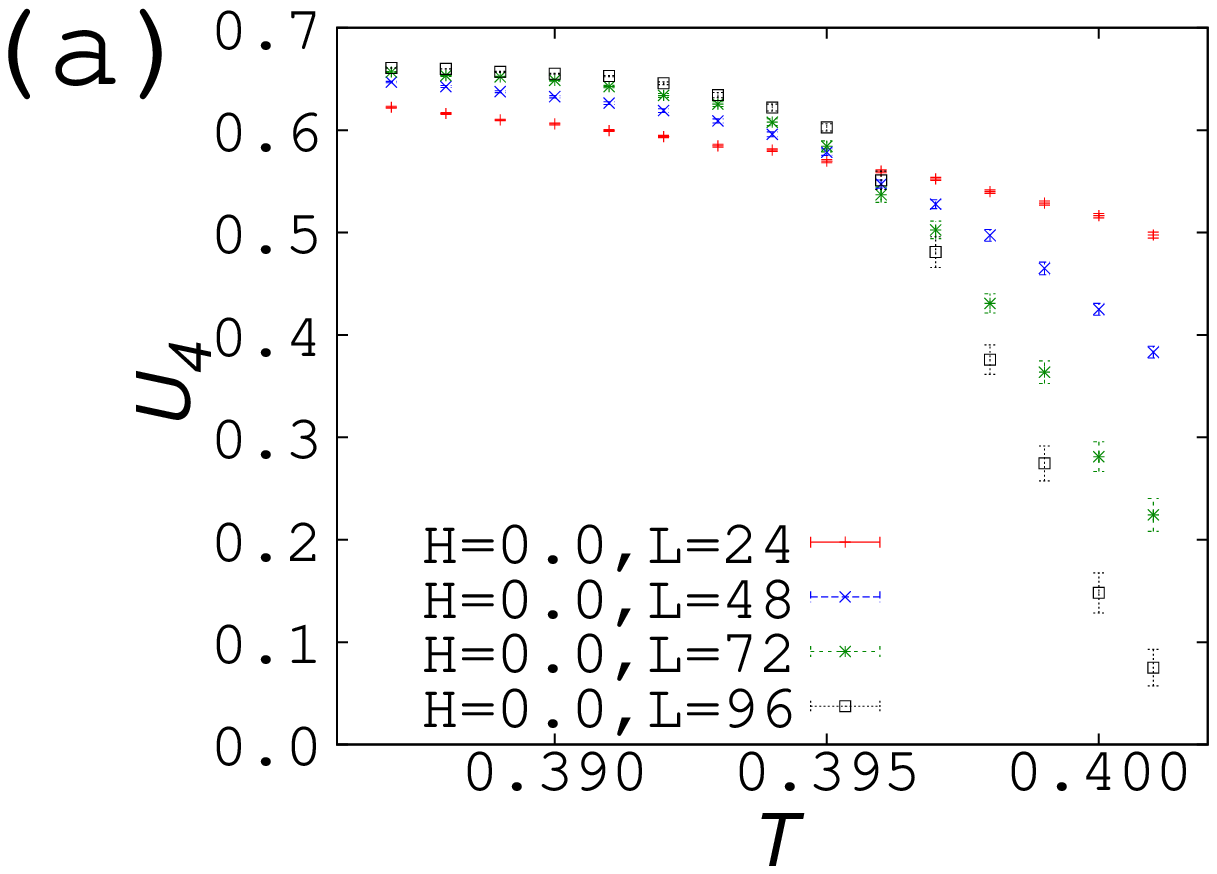}
\includegraphics[width = 7cm]{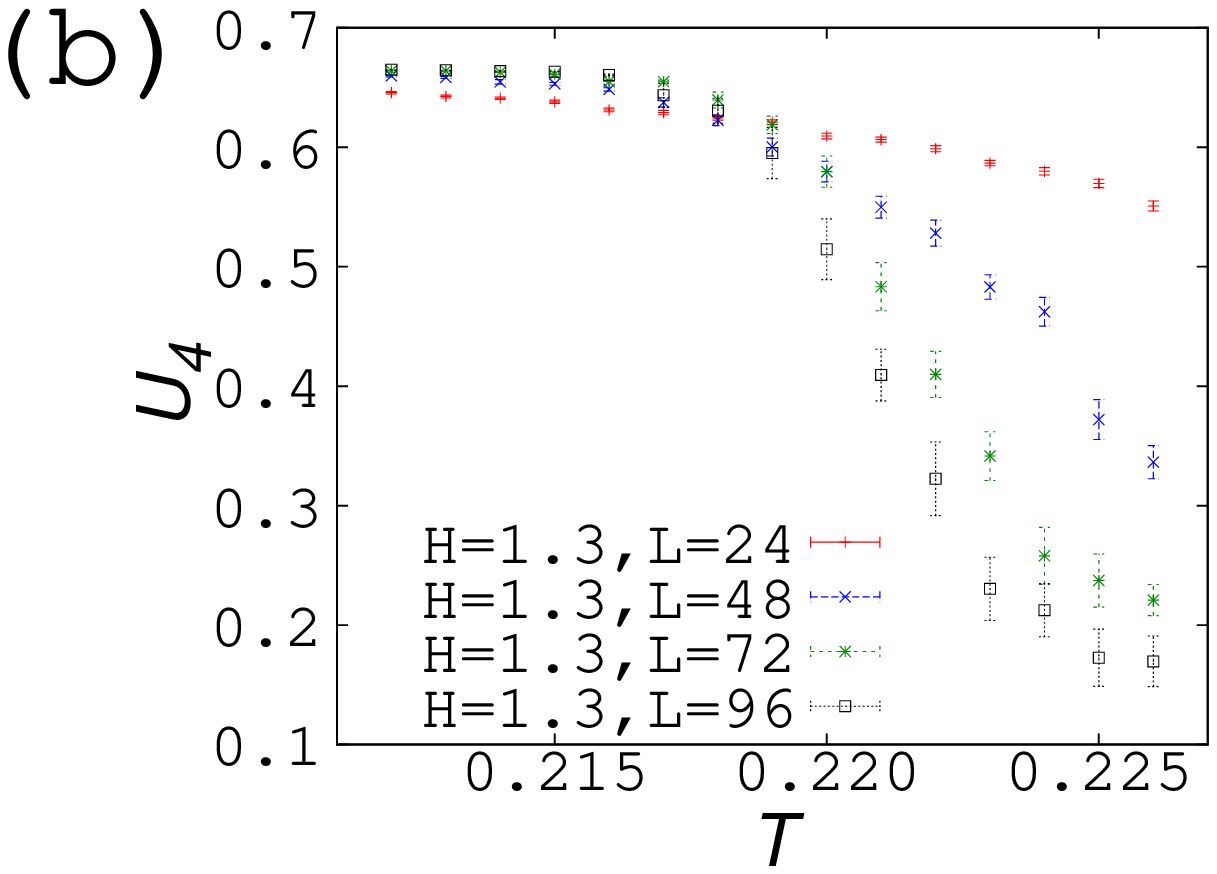}
\end{center}
\caption{Temperature dependence of Binder cumulant at (a) $H=0$ and (b) $H=1.3$.}
\label{Binder_plot}
\end{figure}

\begin{figure}[!tbp]
\begin{center}
\includegraphics[width = 7.5cm]{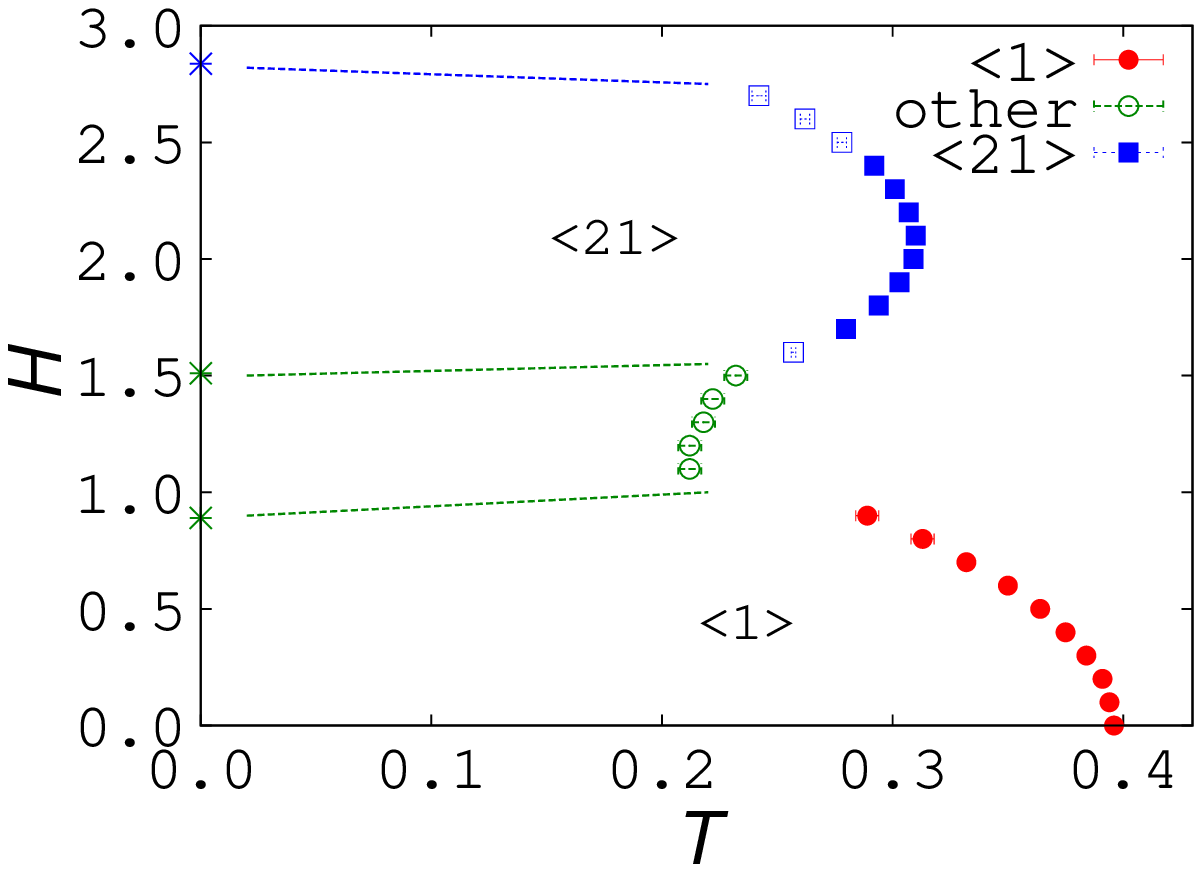}
\end{center}
\caption{Monte Carlo phase diagram.  
The filled circles, squares, and open circles represent the boundaries between the uniform and $\langle 1 \rangle$, $\langle 21 \rangle$, and other complicated phases, respectively. Difference between the filled and open squares is explained in the text. 
Stars stand for the MF phase boundaries of the $\langle 1 \rangle$ and $\langle 21 \rangle$ phases at $T=0$. 
Boundaries between the stripe-ordered phases (dotted lines) are drawn as hand-waving guides for eyes.}
\label{PD_sim}
\end{figure}

We investigate second-order phase transitions by using 
the fourth-order Binder cumulant~\cite{Binder} of $O_{hv}$:
\begin{equation}
U_4 = 1- \frac{ \left< O_{hv} ^4 \right>}{ 3 \left< O_{hv} ^2 \right>}. 
\end{equation}
The transition temperature $T_{\rm c}$ is obtained by the intersection of 
the $U_4(T)$ curves for different system sizes. 
Figs.~\ref{Binder_plot} (a) and (b) depict $U_4(T)$ at (a) $H=0$ and (b) $H=1.3$ for several system sizes, and the $U_4(T)$ curves cross at $T_{\rm c}=0.396(2)$ and $0.218(3)$ for $H=0$ and $1.3$, respectively to exhibit second-order phase transitions. Similar crossings are also observed at all the points denoted by circles in Fig.~\ref{PD_sim} for $H \le 1.5$.
\begin{figure}[b]
\begin{center}
\includegraphics[width = 7.0cm]{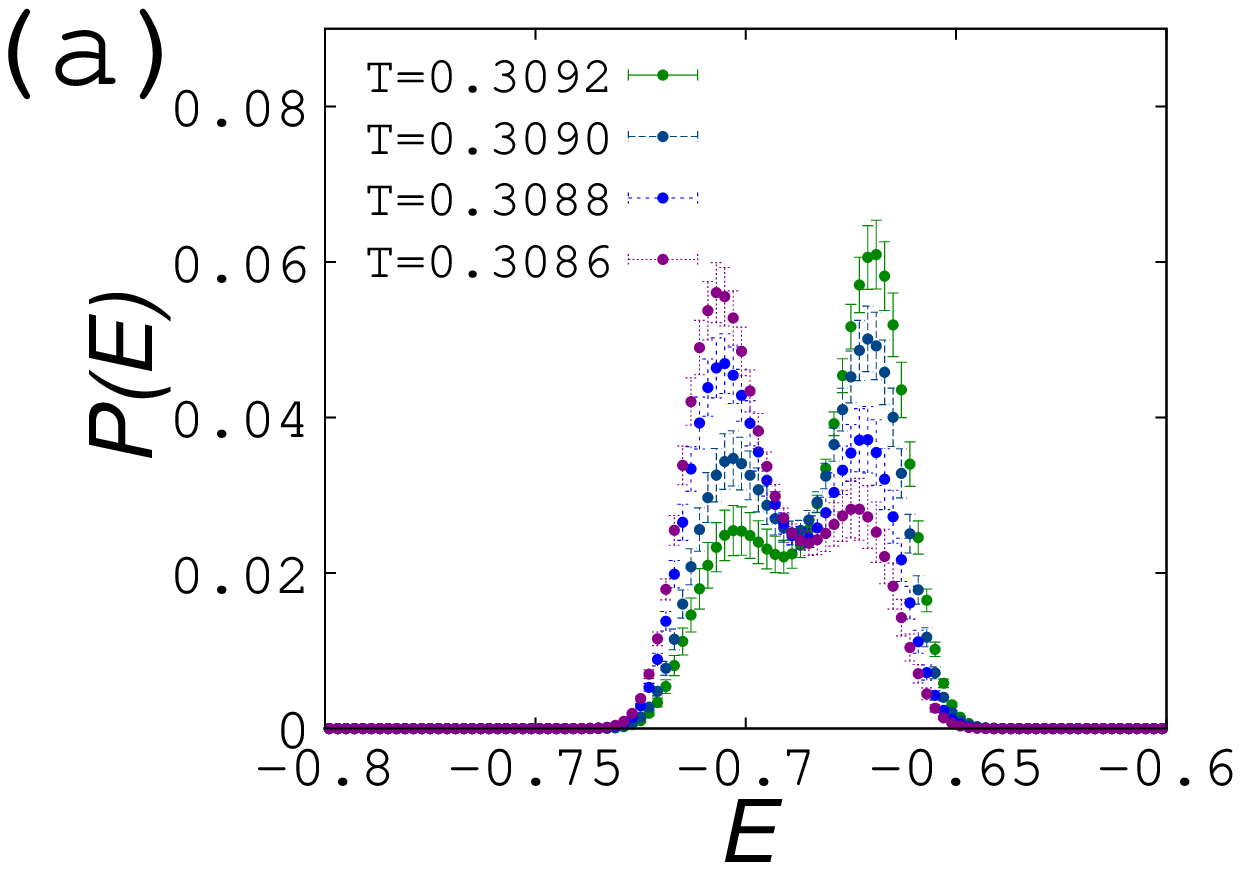}
\includegraphics[width = 7.0cm]{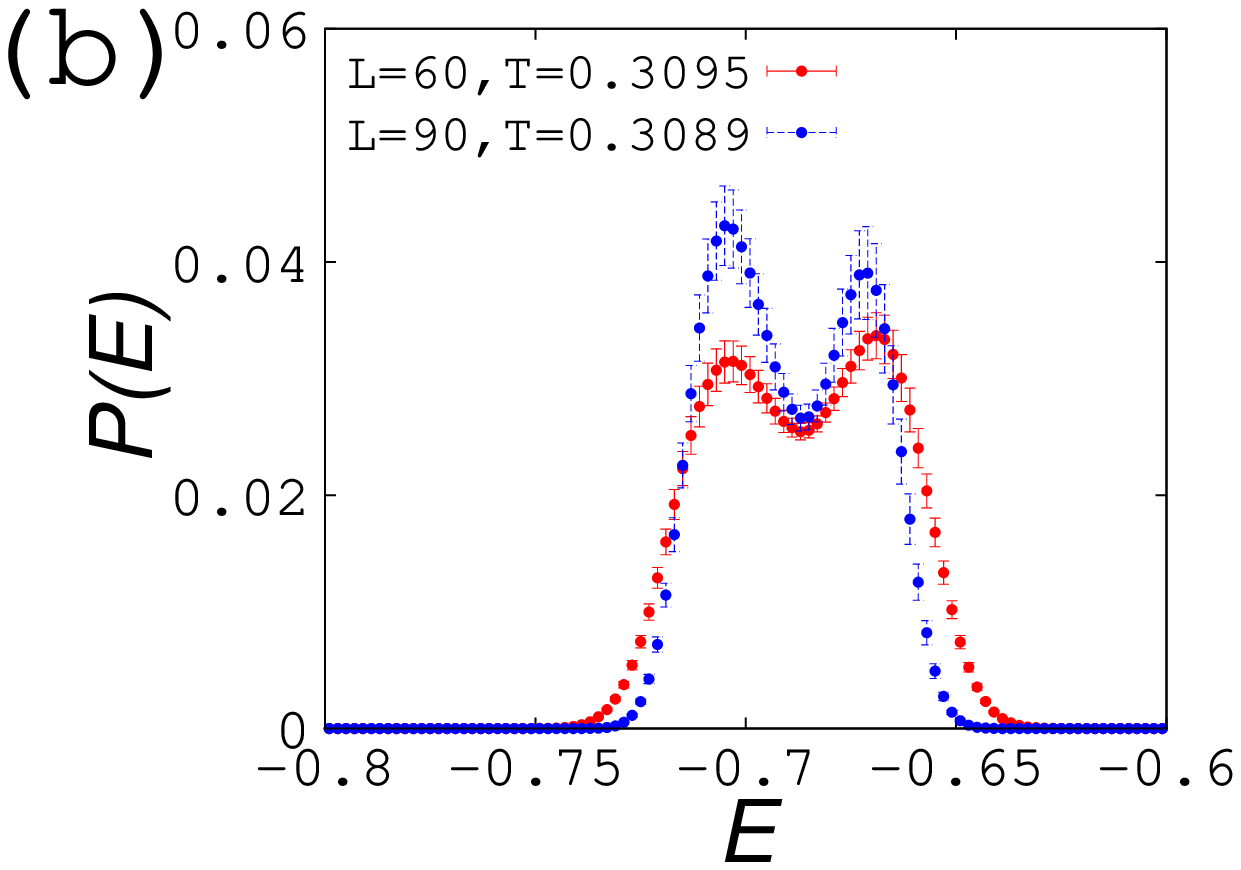}
\end{center}
\caption{Energy histogram at $H = 2.0$ (a) at various temperatures for $L=90$ and 
(b) at $T=0.3095$ for $L=60$ and at $T=0.3089$ for $L=90$ with the double-peak structure.}
\label{HE_h2}
\end{figure}
For first-order transitions, we investigate the histogram of the total energy as a function of $T$ and $H$.  
If the phase transition is of first order, the histogram should have two peaks around the transition temperature, while it should have one peak for second-order transitions. Indeed no double peaks are observed around $T_{\rm c}$ at $H=0$ and $H=1.3$. 
We plot energy histograms around the transition temperature at $H=2.0$ for $L=90$ in Fig.~\ref{HE_h2} (a). 
Double peaks around $E = -0.71$ and $E = -0.67$ are found in the region 
between $T=0.3086$ and 0.3092. 
The lower-energy peak is dominant at $T = 0.3086$, and the higher-energy peak grows and becomes dominant with increasing temperature up to $T = 0.3092$. 
In Fig.~\ref{HE_h2} (b) we compare the double-peak structure for $L=60$ and $90$. The peaks become sharper as the system size increases. 
These observations are clear evidence for the first-order transition. 

To specify the stripe patterns, the structure factor $S(k_x,k_y)$ is investigated: 
\begin{equation}
S(k_x,k_y ) \equiv \left| \tilde{\sigma} _{\vector{k}} \right| ^2 ,
\end{equation}
\begin{equation}
\mathrm{where} \ \  \tilde{\sigma} _{\vector{k}} = \frac{1}{N} \sum _{j} \sigma _j e^{i \vector{k} \cdot \vector{x} _j }. 
\end{equation}
Here $\vector{x} _j$ is the position of site $j$. 
In the present study, the system has $\pi/2$-rotational symmetry and we measure $\bar{S} (k) \equiv S \left( k,0\right) + S \left( 0,k \right) $. 
In Fig.~\ref{Sk} the field dependence of $\bar{S} (k)$ is plotted at $T=0.1$. 
At $H=0.8$, $\bar{S} (k)$ has a peak at $k=\pi$, which indicates the $\langle 1 \rangle$ phase. $\bar{S} (k)$s at $H=1.7$ and $H=2.5$ show peaks at $k=2\pi/3$ and $4\pi/3$, which suggests the $\langle 2 1 \rangle$ phase with the $3$-lattice-unit period. 
We perform the same analyses at different temperatures and fields, and obtain the MC phase diagram displayed in Fig.~\ref{PD_sim}. 

The $\langle 1 \rangle$ phase at low fields and the $\langle 21 \rangle$ phase at high fields are main contribution to the phase diagram, and there exists a region characterized by complicated phases between these two phases, as we see below.

The MF approximation is rigorous at $T=0$ under the assumption of stripe-ordered phases with the $n \leq 11$ restriction, and we substitute the MF phase boundaries of the $\langle 1 \rangle$ and $\langle 21 \rangle$ phases at $T=0$, denoted by stars in Fig.~\ref{PD_sim}, for the MC ones. We sketch rough phase boundaries between the $\langle 1 \rangle$, complicated, and $\langle 21 \rangle$ phases there.

We find that the phase transition between the $\langle 1 \rangle$ and uniform phases is of second order. 
Concerning this phase transition 
at $H=0$, there have been controversial results about the region of $\delta$ for the second-order transition line in the vicinity of $\delta=1$ in the zero-field MC phase diagram~\cite{Pighin,Bab,Horowitz,Fonseca}. 
First-order transition was pointed out at $\delta=1$ in several studies~\cite{Pighin,Rastelli,Cannas}. 
On the other hand, it has been recently shown that the second-order transition line exists up to $\delta \le 1.2585$~\cite{Fonseca,Horowitz,Bab}. Our result at $H=0$ is consistent with the latter, 
and our estimate $T_{\rm c} = 0.396(2)$ is comparable to $T_{\rm c}=0.395(1)$ in Ref.~\cite{Horowitz}.

For the $\langle 2 1 \rangle$ phase at higher fields, we obtain clear evidence for the first-order transition in the energy-histogram analysis for the filled squares in Fig.~\ref{PD_sim}, while we do not observe sharp double peaks for the open squares within the present system sizes. We expect the full first-order phase transition between the $\langle 21 \rangle$ and uniform phases, but tricritical point(s) might exist between the filled and open squares in Fig.\ \ref{PD_sim}. 

We see some complicated phases between the  $\langle 1 \rangle$ and  $\langle 2 1 \rangle$ phases at the intermediate fields. 
The Binder plots for the phase boundaries between these phases and the disordered phase indicate second-order transition as depicted in Fig.~\ref{Binder_plot} (b). At $H=1.3$ $\bar{S} (k)$ has peaks around $4\pi/5$ and $6\pi/5$ (and weakly around $2\pi/5$ and $8\pi/5$), which is consistent with the $\langle 2 1^3\rangle$ phase with the $5$-lattice-unit period. 
It is difficult to identify the patterns of the phases between the $\langle 1 \rangle$ and  $\langle 2 1^3\rangle$ phases and between the $\langle 21 \rangle$ and  $\langle 2 1^3\rangle$ phases because peaks of $\bar{S}_{k}$ are too weak and broad (e.g.\ at $H=1.1$) and positions of them shift continuously as the field changes. 

The MC phase diagram is characterized by the dominant $\langle 1 \rangle$ and $\langle 2 1 \rangle$ phases and the narrow region of complicated phases. This structure brings a field-induced reentrant transition of the $\langle 2 1 \rangle$ phase, i.e., uniform to $\langle 2 1 \rangle$ to uniform phase. 
As an example, the field dependence of the order parameter $O_{hv}$ 
at $T=0.3$ for $L=48$ is shown in Fig.~\ref{Field-dep}.

\section{Summary  \label{summary} }

We investigated the temperature-field phase diagrams for $\delta = 1$
in the two-dimensional dipolar Ising ferromagnet by using the MF and MC methods. In the MC method we adopted the SCO $O(N)$ algorithm, efficient to simulations of long-range interaction systems. 
 In the MC study the finite-size-scaling method was applied for the second-order phase transition and the energy histogram analysis was performed for the first-order phase transitions. We found both qualitative and quantitative differences between  the MF and MC phase diagrams, which is due to the thermal fluctuation and frustration effects. 

The critical temperature at zero field is $T_{\rm c}=0.93$ in the MF analysis, while that in the MC method is $T_{\rm c} = 0.396(2)$. 
The $\langle 31 \rangle$ phase in the MF phase diagram does not 
exist in the MC phase diagram, and the upper transition filed of the $\langle 21 \rangle$ phase is much reduced in the MC phase diagram. 
There is a tricritical point in the phase transition between the $\langle 1 \rangle$ and uniform phases in the MF analysis. On the other hand, this transition is of second order in the MC analysis. 
In the MF phase diagram the $\langle 21 \rangle$ phase neighbors the $\langle 1 \rangle$ phase, and the $\langle 31 \rangle$ phase neighbors the $\langle 21 \rangle$. The $\langle 1 \rangle$, $\langle 21 \rangle$, and $\langle 31 \rangle$ phase boundaries show a nesting structure.

On the other hand, in the MC phase diagram $\langle 1 \rangle$ and $\langle 21 \rangle$ phases do not neighbor and a nesting structure does not exist. Such qualitative differences between the MF and MC phase diagrams have been found in various frustrated systems such as the triangular Ising model and its extension~\cite{Mekata,Landau}. The reduction of $T_{\rm c}$ or $h_{\rm c}$ in the MC method is partially attributed to the thermal fluctuation effects. 

In the MC phase diagram there exist a small region of complicated stripe phases between the regions of $\langle 1 \rangle$ and $\langle 21 \rangle$ phases. 
This feature causes the field-induced reentrant transition of the $\langle 21 \rangle$ phase: uniform to $\langle 2 1 \rangle$ to uniform phase. So far field-induced reentrant transitions have not been observed in the dipolar Ising system such as the case of $\delta=2$~\cite{MM10}. 
We first found a field-induced reentrant transition of a stripe phase, i.e., uniform to $\langle 2 1 \rangle$ to uniform phase. This is a nontrvial characteristic in the model with $\delta=1$.

\section*{Acknowledgments}
The present work was supported by Grants-in-Aid for
Scientific Research C (No. 17K05508) 
from MEXT of Japan, and the Elements Strategy Initiative
Center for Magnetic Materials (ESICMM) under the outsourcing
project of MEXT. Part of numerical calculations were performed on the Numerical 
Materials Simulator at National Institute for Materials Science.

\appendix
\section{Tricritical point in the MF phase diagram}
\label{App1}

We find a tricritical point in the phase transition between the $\langle1\rangle$ and uniform phases. 
Using the translational symmetry $\tilde{U} _{i,j} = \tilde{U} _{i+n , j+n}$, the free energy (\ref{F_MF}) for $n=2$ is rewritten as a function of $m_{\pm} \equiv m_1 \pm m_2$ as follows. 
\begin{eqnarray}
F & =  &  \frac{L^2}{2} \left[ \frac{1}{2} \left\{ \frac{ \tilde{U}_{1,1} + \tilde{U}_{1,2} }{2} \cdot m_+ ^2 +  \frac{ \tilde{U}_{1,1} - \tilde{U}_{1,2} }{2} \cdot m_- ^2 \right\} \right. \nonumber \\
& & - \beta ^{-1} \log \left[ 2 \cosh \left\{ \beta \left( \frac{ \tilde{U}_{1,1} + \tilde{U}_{1,2} }{2} \cdot m_+ + \frac{ \tilde{U}_{1,1} - \tilde{U}_{1,2} }{2} \cdot m_- + H \right) \right\} \right] \nonumber \\
& & \left. - \beta ^{-1} \log \left[ 2 \cosh \left\{ \beta \left( \frac{ \tilde{U}_{1,1} + \tilde{U}_{1,2} }{2} \cdot m_+ - \frac{ \tilde{U}_{1,1} - \tilde{U}_{1,2} }{2} \cdot m_- + H \right) \right\} \right] \right].
\label{F_MF_11}
\end{eqnarray}
In the phase transition from $\langle1\rangle$ to uniform phase, $m_-$ changes 
from $m_- \neq 0$ to $m_- = 0$. 
Hence, the tricritical point is obtained from the following conditions: 
\begin{equation}
\frac{\partial ^2 F}{\partial m_- ^2} \Bigg|_{m_-=0}  = \frac{\partial ^4 F}{\partial m_- ^4} \Bigg|_{m_-=0}  = 0,
\label{TCP1}
\end{equation}
and we have the relations, 
\begin{eqnarray}
&& \frac{ \tilde{U}_{1,1} + \tilde{U}_{1,2} }{2} - \beta \left( \frac{ \tilde{U}_{1,1} + \tilde{U}_{1,2} }{2} \right) ^2 \frac{2}{ \cosh ^2 A }  =  0
\label{TCP2}
\end{eqnarray}
and 
\begin{eqnarray}
&& \frac{4}{ \cosh ^2 A } - \frac{6}{ \cosh ^4 A } =  0 , \label{TCP3}
\end{eqnarray}
\begin{eqnarray}
\mathrm{where} \ \ A = \beta \left( \frac{ \tilde{U}_{1,1} + \tilde{U}_{1,2} }{2} \cdot m_+ + H \right)  . \label{TCPa}
\end{eqnarray}
Solving the SCF equations with the relations, the tricritical point is given as
 $T \simeq 0.6236$ and $H \simeq 3.316$, at which $m_1 = m_2 = \frac{1}{\sqrt{3}} $.

\section{SCO method for the Ising dipolar model}
\label{AppA}

In this appendix we briefly explain the implementation of the SCO algorithm suitable for the Ising dipolar system. 
The SCO method~\cite{SM08} is based on the stochastic potential switching (SPS) algorithm~\cite{Mak05,Mak07} with $O(N)$ switching time, which realizes $O(N)$ computational time for simulation of dipolar systems.   

In the SPS algorithm long-range interaction $V_{ij}$ is stochastically switched to $\tilde{V}_{ij}$ with a probability $P_{ij}$ or to $\bar{V}_{ij}$ with $1-P_{ij}$. 
The potential $\tilde{V}_{ij}$ can be chosen arbitrarily. 
Here $P_{ij}$ is written as 
\begin{equation}
P_{ij} (\sigma _i , \sigma _j )  = \exp \left[ \beta \left( \Delta V_{ij} (\sigma _i , \sigma _j ) - \Delta V_{ij} ^{\ast} \right) \right] ,
\label{SCO_prob1}
\end{equation}
where
\begin{equation}
\ \ \Delta V_{ij} (\sigma _i , \sigma _j )  = V_{ij}(\sigma _i , \sigma _j )  - \tilde{V}_{ij} (\sigma _i , \sigma _j ), 
\end{equation}
and $\Delta V_{ij} ^{\ast}$ is a constant equal to or greater than the maximum value of  $\Delta V_{ij}$. 
The potential $\bar{V}_{ij}$ is given by 
\begin{equation}
\bar{V}_{ij} (\sigma _i , \sigma _j )  = V_{ij} (\sigma _i , \sigma _j )  - \beta^{-1} \log \left( 1-P_{ij} (\sigma _i , \sigma _j )  \right) .
\label{SCO_vbar}
\end{equation}
  In the SCO method for the dipolar Heisenberg model, 
this SPS procedure is applied to all dipolar interactions with $\tilde{V}_{ij}=0$ and is not applied to the nearest-neighbor Heisenberg couplings.
Namely, all nearest-neighbor Heisenberg interactions and 
some dipolar interactions selected by the SPS procedure are considered for the update of the state in one MC step. 
  
\begin{figure}[tbp]
\includegraphics[width = 7.0cm]{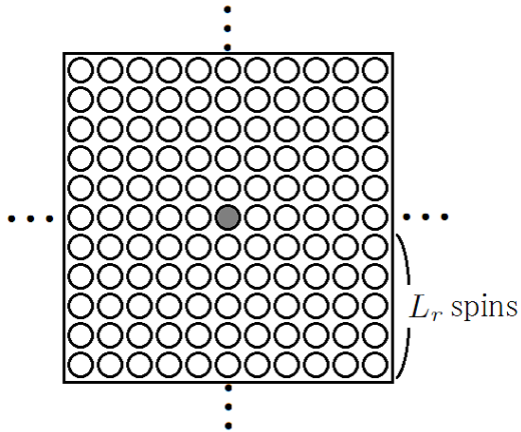}
\caption{ The range we do not use the SCO method: In the MC simulation of this study, we update the state with naive MC method if vertical and horizontal distance between one spin (gray circle) of a bond and the other is less than 5 (inside the square), and adopt the SCO method otherwise (outside the large square). }
\label{SCO_range}
\end{figure}

Here we tune the number of the dipolar interactions to which the SPS procedure is applied. We show that this tuning is useful for the acceleration of the relaxation in the Ising dipolar model. 
For a single spin flip for the spin denoted by the filled circle in Fig.~\ref{SCO_range}, the SPS procedure is not applied to the dipolar interactions around the spin within the range of $(2L_r+1) \times (2L_r+1)$ sites (open circles in Fig.~\ref{SCO_range}) , i.e., these interactions are always counted for the single spin flip in the same way as the nearest-neighbor Ising interactions. 

\begin{figure}[tbp]
\begin{center}
\includegraphics[width = 9.0cm]{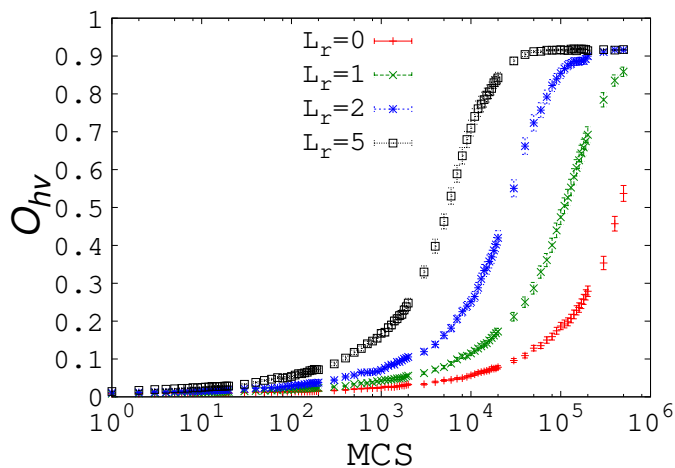}
\end{center}
\caption{$L_r$ dependence of relaxation curves of $O_{h \nu}$ at $T=0.35$ with $\alpha=1.5$ for $L=96$.}
\label{SCO_relax}
\end{figure}

We give $L_r$ dependence of the relaxation curves of $O_{h\nu}$ in Fig.~\ref{SCO_relax}. 
The case $L_r=0$ corresponds to the original SCO method and we find drastic reduction of the relaxation time for $L_r \ne 0$. We adopted $L_r=5$, which provides enough efficiency in the MC sampling for the present work. 

This improvement is attributed to the discretized spin state (up or down) of the Ising spin. 
The dipolar interaction in eq.~(\ref{Hamiltonian}), 
\begin{equation}
 V_{ij} (\sigma _i , \sigma _j )  = \frac{\sigma _i  \sigma _j}{r^3}, 
\end{equation}
takes two values, i.e., 
$V_{ij} (\sigma _i , \sigma _j )=\frac{1}{r^3}$ and $-\frac{1}{r^3}$ 
for ferromagnetic ($\sigma _i=\sigma _j= \pm1$) and antiferromagnetic ($\sigma _i=-\sigma _j= \pm1$) spins, respectively. 
Defining $\Delta V_{ij} ^{\ast}=\alpha \frac{1}{r^3}$, in which $\alpha>1$, $P_{ij} (\sigma _i , \sigma _j )  = \exp(\frac{1-\alpha}{r^3T})$ and 
$\exp(\frac{-1-\alpha}{r^3T})$ for ferromagnetic and antiferromagnetic spins, respectively. In the present study we take $\alpha=1.5$. 

Because
\begin{equation}
\frac{P_{ij} (\sigma _i=1 , \sigma _j=1 )}{ P_{ij} (\sigma _i=1, \sigma _j=-1)}  = \exp \Big(\frac{2}{r^3T}\Big), 
\end{equation}
$P_{ij}$ for ferromagnetic spins is much larger than that for antiferromagnetic spins for smaller $r$ and lower $T$. 
Namely the probability for selecting $\bar{V}_{ij}$ for ferromagnetic spins is much smaller than that for antiferromagnetic spins. 
This large difference of $P_{ij} (\sigma _i, \sigma _j)$ causes a deviation of the potential switching pattern and inefficiency of the MC sampling.  

\begin{thebibliography}{99}


\bibitem{Bader}
S.~D.~Bader, Rev.\ Mod.\ Phys \textbf{78}, 1 (2006).

\bibitem{ASB90} R.~Allenspach, M.~Stampanoni, and A.~Bischof, 
Phys.\ Rev.\ Lett.\ \textbf{65}, 3344 (1990).

\bibitem{Kashuba} A.~B.~Kashuba and V.~L.~Pokrovsky, 
Phys.\ Rev.\ B \textbf{48}, 10335 (1993).

\bibitem{DeBell} K.~De'Bell, A.~B.~MacIsaac, and J.~P.~Whitehead, 
Rev.\ Mod.\ Phys.\ \textbf{72}, 225 (2000). 

\bibitem{Portmann} O.~Portmann, A.~Vaterlaus, and D.~Pescia, Nature \textbf{422}, 701 (2003).






\bibitem{Booth} I.~Booth, A.~B.~MacIsaac, J.~P.~Whitehead, and K.~De'Bell, 
Phys.\ Rev.\ Lett.\ \textbf{75}, 950 (1995).



\bibitem{MacIsaac}  A.~B.~MacIsaac, J.~P.~Whitehead, M.~C.~Robinson, 
and K.~De'Bell, Phys.\ Rev.\ B {\textbf 51}, 16033 (1995).

\bibitem{Toloza} 
J.~H.~Toloza, F.~A.~Tamarit, and S.~A.~Cannas, Phys.\ Rev.\ B \textbf{58}, R8885 (1998).

\bibitem{Rastelli06} E.~Rastelli, S.~Regina, and A.~Tassi, Phys.\ Rev.\ B \textbf{73}, 144418 (2006).

\bibitem{Cannas} S.~A.~Cannas, M.~F.~Michelon, D.~A.~Stariolo, and F.~A.~Tamarit, 
Phys.\ Rev.\ B \textbf{73}, 184425 (2006).

\bibitem{Pighin} S.~A.~Pighin and S.~A.~Cannas Phys.\ Rev.\ B \textbf{75}, 224433 (2007). 

\bibitem{Rastelli} E.~Rastelli, S.~Regina, and A.~Tassi, Phys.\ Rev.\ B \textbf{76}, 054438 (2007).

\bibitem{Vindigni} A.~Vindigni, N.~Saratz, O.~Portmann, D.~Pescia, and P.~Politi, 
Phys.\ Rev.\ B \textbf{77}, 092414 (2008).

\bibitem{Rizzi} L.~G.~Rizzi and N.~A.~Alves, Physica B \textbf{405}, 1571 (2010).

\bibitem{Fonseca} J.~S.~M.~Fonseca, L.~G.~Rizzi, and N.~A.~Alves, 
Phys.\ Rev.\ E \textbf{86}, 011103 (2012).

\bibitem{Ruger} R.~R\"{u}ger and R.~Valent\'{i}, Phys.\ Rev.\ B \textbf{86}, 024431 (2012). 

\bibitem{Horowitz} C.~M.~Horowitz, M.~A.~Bab, M.~Mazzini, M.~L.~Rubio Puzzo, 
and G.~P.~Saracco, Phys.\ Rev.\ E \textbf{92}, 042127 (2015).

\bibitem{Bab} M.~A.~Bab, C.~M.~Horowitz, M.~L.~Rubio Puzzo, and G.~P.~Saracco, 
Phys.\ Rev.\ E \textbf{94}, 042104 (2016).



\bibitem{Abanov} Ar.~Abanov, V.~Kalatsky, V.~L.~Pokrovsky, and W.~M.~Saslow, 
Phys.\ Rev.\ B \textbf{51}, 1023 (1995).

\bibitem{Garel} T.~Garel and S.~Doniach, Phys.\ Rev.\ B \textbf{26}, 325 (1982).
\bibitem{Arlett} J. Arlett J. P. Whitehead, A. B. MacIsaac, and K. De'Bell, 
Phys.\ Rev.\ B \textbf{54}, 3394 (1995).

\bibitem{MM10} R.~D\'{i}az-Mendez and R.~Mulet, Phys.\ Rev.\ B \textbf{81}, 184420 (2010).

\bibitem{Berezinskii} V.~L.~Berezinskii, Sov.\ Phys.\ JETP {\textbf 32}, 493 (1971); 
{\textbf 34}, 610 (1972).

\bibitem{KT} J.~M.~Kosterlitz and D.~J.~Thouless, J.\ Phys.\ C {\textbf 6}, 1181 (1973).




\bibitem{SM08} M.~Sasaki and F.~Matsubara, J.\ Phys.\ Soc.\ Jpn.\ \textbf{77}, 024004 (2008). 

\bibitem{FT09} K.~Fukui and S.~Todo, J.\ Comput.\ Phys.\ {\textbf 228}, 2629 (2009).

\bibitem{H18} T.~Hinokihara, M.~Nishino, Y.~Toga, and S.~Miyashita, 
Phys.\ Rev.\ B \textbf{97}, 104427 (2018). 

\bibitem{Binder} 
K.~Binder, Phys.\ Rev.\ Lett.\ {\textbf 47}, 693 (1981).


\bibitem{Mekata} M.~Mekata, J.\ Phys.\ Soc.\ Jpn.\ {\textbf 42}, 76 (1977).

\bibitem{Landau}
D.~P.~Landau, Phys.\ Rev.\ B {\textbf 27}, 5604 (1983).


\bibitem{Mak05} C.~H.~Mak, J.\ Chem.\ Phys.\ \textbf{122}, 214110 (2005).

\bibitem{Mak07} C.~H.~Mak and A.~K.~Sharma, Phys.\ Rev.\ Lett.\ \textbf{98}, 180602 (2007). 





\end{thebibliography}
\end{document}